# Interpreting "altmetrics": viewing acts on social media through the lens of citation and social theories

Stefanie Haustein[*,1], Timothy D. Bowman[1] & Rodrigo Costas[2]

[*]*stefanie.haustein@umontreal.ca*

[1] École de bibliothéconomie et des sciences de l'information, Université de Montréal, C.P. 6128, Succ. Centre-Ville, Montréal, QC, H3C 3J7 (Canada)

[2] Center for Science and Technology Studies, Leiden University, Wassenaarseweg 62A, 2333 AL Leiden (The Netherlands)

## 1. Introduction

More than 30 years after Blaise Cronin's seminal paper (Cronin, 1981; see reprint in this book) the metrics community is once again in need of a new theory, this time one for so-called "altmetrics". Altmetrics, short for alternative (to citation) metrics —and as such a misnomer— refers to a new group of metrics based (largely) on social media events relating to scholarly communication. The term originated on 29 September 2010 in a tweet by Jason Priem in which he uttered his preference for the word *altmetrics* in the context of various metrics provided for PLOS journal articles: "I like the term #articlelevelmetrics, but it fails to imply *diversity* of measures. Lately, I'm liking #altmetrics." (Priem, 2010). Although Priem is responsible for coining the term, the idea of measuring broader scientific impact through the web and had been discussed by Cronin and others (e.g., Almind & Ingwersen, 1997; Cronin, Snyder, Rosenbaum, Martinson, & Callahan, 1998; Cronin, 2001; see also Thelwall's chapter in this book) in the context of webometrics years before:

> Scholars may be cited formally, or merely mentioned en passant in listservs and others electronic discussion fora, or they may find that they have been included in reading lists or electronic syllabi. Polymorphous mentioning is likely to become a defining feature of Web-based scholarly communication. (Cronin et al., 1998)

> There will soon be a critical mass of web-based digital objects and usage statistics on which to model scholars' communication behaviors—publishing, posting, blogging, scanning, reading, downloading, glossing, linking, citing, recommending, acknowledging—and with which to track their scholarly influence and impact, broadly conceived and broadly felt. (Cronin, 2005b)

Priem —co-author of the altmetrics manifesto (Priem, Taraborelli, Groth, & Neylon, 2010) and co-founder of ImpactStory[1], an online tool aggregating various metrics on the researcher level— and colleagues argued that metrics based on 'traces' of use and production of scholarly output on social media platforms could help to improve scholarly communication and research evaluation. The term *altmetrics* was introduced out of the need to differentiate these new metrics from traditional citation-based indicators, which the altmetrics movement is seeking to replace or use as an alternative. The altmetrics manifesto and other work by Priem and colleagues appeal to the scientific community and research managers to "value all research products" (Piwowar, 2013), not just journal articles, and to measure impact in a broader sense by looking at more than just citations. The manifesto lists various sources of new metrics that would complement and replace traditional forms of publication, peer review, and citation analysis (Priem et al., 2010). Priem (2014) claimed that with scholarship moving online, former invisible aspects of scholarly communication—such as reading, discussing and recommending scientific papers—leave traces that can be collected earlier and easier than citations and would thus provide an alternative to citations. The idea of altmetrics resonated with some in the scholarly community, academic libraries, publishers, and (particularly) with research managers and funders, who were attracted by the idea of measuring the impact of research on the broader non-scientific community as a return on their investment (Adie, 2014). Within the bibliometric community the examination of social bookmarking data as indicators of readership was the first examination of altmetrics (Haustein & Siebenlist, 2011). Bornmann (2014; see also his chapter in this book) even argued that scientometrics was undergoing a scientific revolution (Kuhn, 1962) due to taxonomy changes regarding the definition of impact (i.e., from scientific to a broader concept of impact).

---
[1] https://impactstory.org/

Although hopes are high that these new metrics are able to capture research impact earlier or more broadly than citations (compare Bornmann's, Moed's and Thelwall's chapters in this book), they are limited by the technological ecosystems from which they are captured as they seem to measure what is technically feasible instead of what is sensible (Taylor, 2014). While there is a debate among scholars on how to properly define these new metrics, they are considered by some universities, libraries, publishers, funders and other organizations to evaluate and assess scholarly output. In the context of the Research Excellence Framework (REF), the Higher Education Funding Council for England (HEFCE) decided to "include all kinds of social, economic and cultural benefits and impacts beyond academia" (HEFCE, 2011, p. 4). This has introduced a certain level of social pressure for scholars to understand, participate in, and manage their use of computer-mediated environments, as there is the possibility that their events within these contexts will be recorded and made available to others for evaluation. These new metrics are supposed to provide insight into the measure of societal impact, as they are able to track events outside the scientific community revolving around scholarly output (Priem et al., 2010; Priem, 2014).

Just as there is a need for citation theory, there is also a need to define the meaning of the various indicators grouped under the term altmetrics. As depicted in Cronin (1981), the search for the meaning of citations was pervasive in the early days of citation analysis. Similar to Gilbert (1977, p. 114), who stated that "we do not yet have any clear idea about what exactly we are measuring when we analyze citation data", the question "What do they actually mean?" has been a reoccurring one in research on altmetrics. Several parallels can be drawn between the early days of citation analysis and today's search for the meaning and need for a theoretical framework of social media metrics. What is considerably different, however, is that altmetrics capture events on platforms that are in a constant state of change and whose use and user communities are new, diverse and not entirely understood, while the act of citing (although not always being counted) has existed since the early days of modern science. While social rules and norms exist within the scientific community of how, when, and what to cite, these norms are not yet established in social media as the ecosystem is in a state of flux.

## 2. Defining and classifying social media events and metrics

Although altmetrics are generally understood as metrics that measure impact beyond citations and define scholarly output in a broader sense than only peer-reviewed journal articles, there is no common and agreed-upon definition or understanding of altmetrics except for that they "capture very different things" (Lin & Fenner, 2013, p. 20). The only unifying concept is that they stand in opposition to 'traditional' bibliometrics and common practices in research evaluation, especially considering citations (Priem, 2014). Altmetrics even include download and article usage statistics, although these have been available much longer than social media applications (Borghuis, 1997; Kaplan & Nelson, 2000; Luther, 2001). Priem (2014, p. 266) defines altmetrics as the "study and use of scholarly impact measures based on activity in online tools and environments" and as such as a proper subset of webometrics (compare Thelwall's chapter in this book). As a pragmatic attempt at a suitable definition for altmetrics, one could say that these metrics are: *events on social and mainstream media platforms related to scholarly content or scholars, which can be easily harvested (i.e., through APIs), and are not the same as the more 'traditional' concept of citations*.

The criticism for the term *alt*metrics has grown as more empirical studies have found that most social media based indicators are (if at all) complements and not alternatives to citation-based indicators. Rousseau and Ye (2013, p. 2) stated that altmetrics was "a good idea but a bad name" and proposed *influmetrics*—initially introduced by Elisabeth Davenport and discussed by Cronin and Weaver (Cronin & Weaver, 1995) in the context of acknowledgements and webometrics (Cronin, 2001)—as an alternative term to "suggest diffuse and often imperceptible traces of scholarly influence – to capture the opportunities for measurement and evaluation afforded by the new environment" (Cronin, 2005b). Haustein and colleagues (Haustein, Larivière, Thelwall, Amyot, & Peters, 2014) used *social media metrics* to emphasize from which data sources the metrics were obtained without attempting to describe intent or meaning and to better differentiate new indicators from more traditional ones (i.e., citation and downloads). Although *social media metrics* seems a better fit as an umbrella term because it addresses the social media ecosystem from which they are captured, it fails to incorporate the sources that are not obtained from social media platforms (such as mainstream newspaper articles or policy documents) that are collected (for instance) by Altmetric.com. As current definitions of altmetrics are shaped and limited by active platforms, technical possibilities, and business models of aggregators such as Altmetric.com, ImpactStory, PLOS, and Plum Analytics—and as such constantly changing—this work refrains from defining an umbrella term for these very heterogeneous new metrics.

Instead a framework is presented that describes *acts* leading to (online) events on which the metrics are based. These acts refer to activities occurring in the context of social media, such as discussing on Twitter or saving to Mendeley, as well as downloading and citing. The framework groups various types of acts into three categories—accessing, appraising, and applying—and provides examples of actions that lead to visibility and traceability online. These are the *polymorphous mentions* Cronin and colleagues (1998, p. 1320) anticipated. In order to discuss the traces that these acts leave online, the following generic terms — as agreed upon at the 2014 PLOS ALM workshop (Bilder, Fenner, Lin, & Neylon, 2015)— have been adopted:

- **research object:** a scholarly object, for which an event can be recorded;
- **event**: a recorded activity or action which relates to the research object;
- **host**: the place where research objects are made available and exposed to potential events;
- **source**: a platform where events are available;
- **consumer**: a party that collects and uses events to research objects
    - *aggregator:* a type of consumer who collects and provides events to research objects with a specific methodology;

- *end user* or *audience:* a type of consumer who uses and applies events in a specific context and intention.

As acts and recorded events will differ whether they focus on, for example, a journal article or a researcher, this framework distinguishes between scholarly agents and scholarly documents as two particular categories of research objects. Agents (Bourdieu, 1975) include individual scholars, research groups, departments, universities, funding organizations and others entities acting within the scholarly community. Following Otlet's (1934, p. 6) broad definition of a document as "a set of facts or ideas presented in the form of a text or image"[2], this category includes traditional scholarly publications (e.g., journal articles, book chapters, conference proceedings, monographs, theses, reports and other types of grey literature), patents, presentations and lectures, as well as blog posts, datasets, software code, and other forms of scholarly work and output. This dichotomy between agent and document allows one to consider that 'altmetrics' can appear not only as article-level metrics[3], but that it can also be applied to a broad spectrum of research objects.

In order to differentiate between various acts leading to online events on different sources in relation to the document or agent, we propose a framework that classifies these acts into three categories (Figure 1). We argue that these three categories—*access*, *appraise* and *apply*—capture various stages and facets of use and interactions with research objects. The framework is designed to incorporate all main act types leading to events related to scholarly documents and agents. Although it does not claim to be exhaustive as to include all types of possible events—particularly in terms of future changes regarding technology and affordance[4] use—but it is assumed that the categories should be broad enough to incorporate new developments when required. A framework is proposed in this instance because it allows one to consider the "system of concepts, assumptions, expectations, beliefs, and theories that supports and informs" the problem (Maxwell, 2009, p. 222). It is thought to improve the understanding of the various heterogeneous acts that relate to different research objects.

As shown in Figure 1, each of the three act categories, *access*, *appraise* and *apply*, includes various types of acts, which differ slightly depending on the research object. For example, *applying* a document would comprise reusing and building upon theories, software, or datasets, while for an agent the act of *applying* refers to collaboration. Emphasized by the spiral layout, it is generally assumed that the level of engagement increases as one moves across categories of acts from *accessing* over *appraising* to *applying* (i.e. inwards across layers in Figure 1), as well as across types of acts within categories (i.e. clock-wise along the spiral). For example, acts related to engagement with a journal article increases within the *access* category as one moves from viewing a paper title to storing it in a reference manager or within the *appraise* category as one moves from a quick mention on Twitter to a mention in a policy document. It is also important to note that the boundaries between specific categories and types are fuzzy, as they can vary and overlap based on uses or contexts.

---

[2] Translated by the authors from: "un ensemble de faits ou d'idées présentés sous forme de texte ou d'image" (Otlet, 1934, p. 6)

[3] It should also be noted that from an indicator perspective, events related to a particular research object can also be aggregated. For example, if the act of saving a journal article to Mendeley leads to a recorded number of Mendeley readers, these event counts can be aggregated for all documents of a particular agent associated with the documents (e.g., author, journal, discipline, country). However, even though the indicator refers to the agent, the recorded event relates to the research object as the smallest level of analysis, in this case the document and not the particular agent.

[4] Affordances are observed qualities of an object within a context that allow for some type of action (Gibson, 1977).

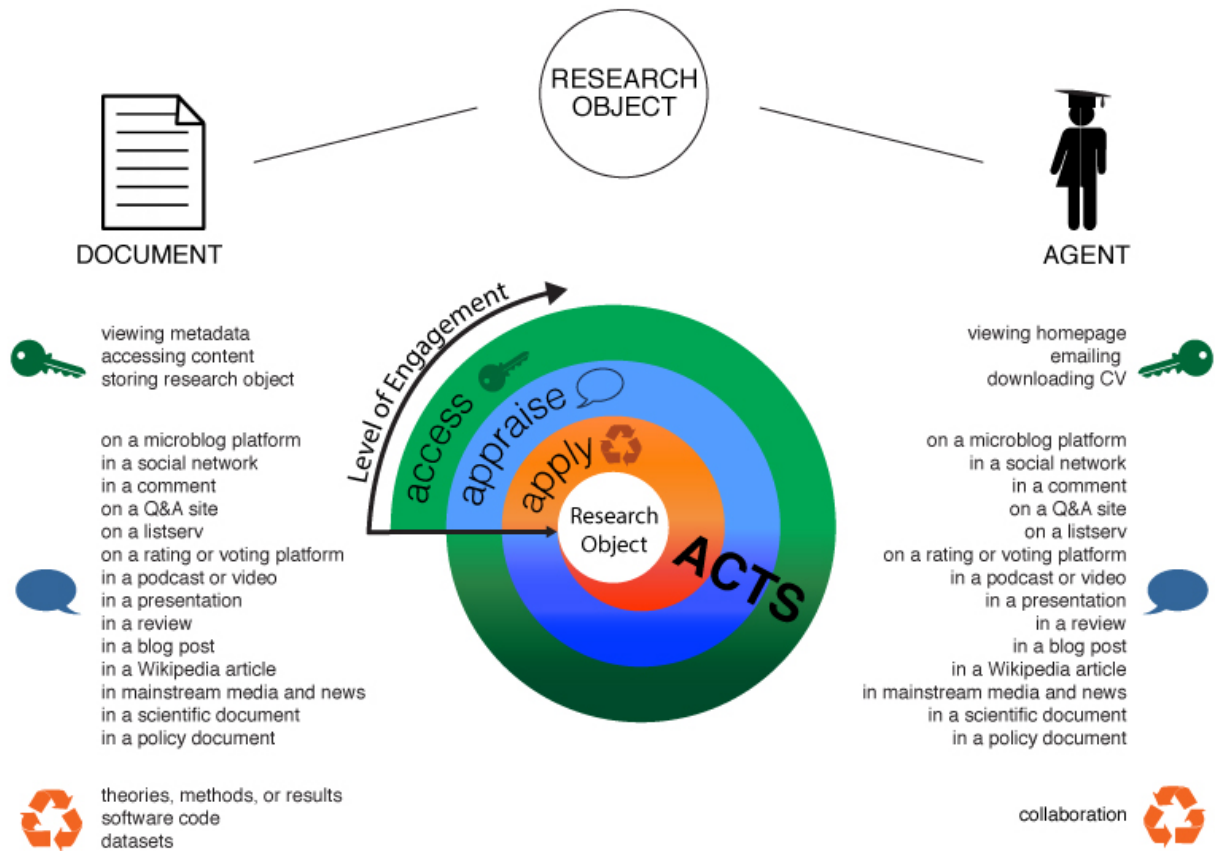

**Figure 1.** Framework of categories and types of acts referring to research objects (scholarly documents and agents).

*Access*. This category refers to acts that involve accessing and showing interest in the research object. In the case of scholarly documents this includes *viewing metadata*, which refers to viewing the title, abstract, or description of, for example, a paper or book, presentation slides, datasets, or software. *Accessing content* includes viewing and downloading the entire document, while *storing the research object* implies making it available for future use. Online events which currently capture these acts include view and download counts on various platforms and repositories (e.g., journal websites, Dryad, FigShare, SlideShare, Github) and reader and bookmarking counts on reference managers such as BibSonomy, CiteULike, Mendeley and Zotero.

Focusing on agents as research objects, the access category includes, for example, *viewing* a university's or scholar's homepage or user profile on platforms such as ResearchGate or Academia.edu, *accessing* the agent through electronic means (e.g. email, messaging, Skype, etc.), and *storing* their information for future use, for example by downloading a scholar's CV or 'friending' or 'following' them on a social media platform such as Twitter, ResearchGate, or Academia.edu.

*Appraise*. The category of appraising includes the act of mentioning the research object on various platforms such as a microblogs, in a social network, in a comment, on a Q&A site, listserv, or rating or voting platform, as well as in a podcast or video, presentation, review, blog post, Wikipedia article, mainstream media and news, scientific or policy document. These appraisal acts are almost identical for both types of research objects except for particular technical differences and affordance use, which can be different for agents and documents. For example, mentioning a scholarly paper on Twitter usually implies linking to it via a URL or document identifier (e.g., http://dx.doi.org/10.1038/504211a), while mentioning a researcher implies using the '@' symbol followed by a Twitter handle (e.g., @csugimoto to mention Cassidy R. Sugimoto).[5] With increasing level of engagement, appraising a scholarly document or agent can range from a brief mention in a post on Twitter or Facebook to a citation in a policy document. Various acts of appraisal can be expressed in a comment, on rating and voting systems (which are usually crowd-sourced and quantitative) such as rating functions on Reddit, Upworthy or RateMyProfessor.com, or as brief discussions and more extensive and qualitative (peer) judgment, typically carried out by an expert (e.g., on F1000, Pubpeer and ResearchGate).

*Apply*. In terms of scholarly documents, we define apply as actively using significant parts of, adapting, or transforming the research object. This occurs in the form of applying theories, frameworks, methods, or results from a scientific document, software code, or dataset(s) as a foundation to create new work. In scholarly documents, applying the content of other documents (and to a lesser extent datasets and software code) is usually indicated through a citation, in which case the distinction between the event categories 'mentioning' and 'applying' may become blurry. However, applying suggests a much higher degree of engagement with the original content than is found in the access or appraise categories. Examples of types of applying acts include the thorough discussion of an article's content in a blog, the use of a scholarly document for self-study, the adaptation of the content of an article for a lecture, the modification or improvement of a dataset or software, or even the use of scholarly output for commercial purposes. Regarding agents, the apply category refers to the act of collaboration. The scholar's knowledge, experience, and reputation are used to formulate something that did not exist before. It may also refer to the participation of the scholar in Q&A sites (such as ResearchGate) where their involvement helps to answer questions.

## 3. Introducing potentially relevant theories

To improve the understanding of the acts resulting in online events from which metrics are collected, select citation and social theories are used below to interpret the phenomena being measured. Citation theories are used because the new metrics based on these events are supposed to replace or complement citations as indicators of impact. Social theories, on the other hand, are discussed because there is an inherent social aspect to the measurements and because scholars may face pressures to ensure their work has societal impact (HEFCE, 2011).

### 3.1. Citation theories

Knowledge about citing behavior and the symbolic characteristics of citations has always been considered essential to determine whether it makes sense to use citation analysis in various areas of application, particular in the context of impact metrics and research evaluation. However, a complete theory of citation is lacking (Cronin, 1981; Leydesdorff, 1998). The increasing use of social media in scholarly communication comes with the same demand, as theories and frameworks are required to assess the meaning of, and to validate, new indicators as performance and impact metrics (Wouters & Costas, 2012). Citation theories discussed here are the *normative theory*, the *social constructivist theory*, and *concept*

---

[5] Of course, mentions of scholars, as well as scholarly output, occur on Twitter without the use of these particular affordances (e.g., "Cassidy R. Sugimoto", "Sugimoto's latest paper in JASIST"), but they are not recognized as such by Twitter and their proper identification requires more sophisticated tools.

*symbols*.[6] The normative and social constructivist approaches can be considered as two of the most important (and opposing) facets of citation theory (Cronin, 2005) that are still discussed and tested today (Riviera, 2014). In addition, Small's (1978) concept symbol theory has been intensively discussed in the literature, particularly in the context of 'obliteration by incorporation' (Merton, 1968a) and for the study of the 'socio-cognitive location' of scholars (Costas & Van Leeuwen, 2012; Moed, 2005), and has been considered in recent conceptual discussions in the field of Scientometrics (Guns, 2013).

### 3.1.1. Normative theory

According to the normative theory, citations are indirect indicators of intellectual influence, reflecting norms and values of science through which scholars are expected to acknowledge the use of the cited work (Kaplan, 1965). Merton (1973) defined the *ethos of science* (i.e., the set of norms and values that rule science) in terms of the four basic norms: communism, universalism, disinterestedness, and organized skepticism. As the basis of normative citation theory, Merton's "sociology of science provides the most coherent theoretical framework available" (Small, 2004, p.72). Although the normative citation theory is based on the assumption that referencing behavior is guided by these norms, it does not claim that authors always strictly adhere to it (De Bellis, 2009; Moed, 2005). In Merton's words the *communism norm* refers to the "nontechnical and extended sense of common ownership of goods" (Merton, 1973, p. 273). Particularly in the context of citations, the well-known idea of "giving credit where credit is due" is attributed to this norm, as authors acknowledge the value of a colleague's work by citing it. *Universalism*, as defined by Merton, "finds immediate expression in the canon that truth-claims, whatever their source, are to be subjected to pre-established impersonal criteria" (Merton, 1973, p. 210). Thus, this norm ascertains that all scientists can contribute to science and are expected to evaluate the works of others regardless of non-scientific characteristics such as race, nationality, culture, or gender. Merton (1988, p. 621) argued that "symbolically, [the reference] registers in the enduring archives the intellectual property of the acknowledged source by providing a pellet of peer recognition of the knowledge claim". Thus, according to the normative theory, citations are the rewards in the science system indicating fair cognitive and intellectual influence.

Scientists are supposed to act for the benefit of a common scientific enterprise, rather than professional gain, as expressed by the *disinterestedness norm*. In Merton's (1973, p. 276) words "a passion for knowledge, idle curiosity, altruistic concern with the benefit to humanity, and a host of other special motives have been attributed to the scientist". In the context of citation analysis, Nicolaisen (2007, p. 617) argues that "it assumes that scientists are disinterested and do not seek to gain personal advantages by flattering others or citing themselves". According to the *organized skepticism norm*, scientific claims must be exposed to critical scrutiny before being accepted. From a citation analysis perspective this is directly related with the publication process of scientific results and new knowledge, as scientists must treat any new claim with skepticism, including their own contributions. Frequently, the *norm of originality* is also included (Ziman, 2000) among the Mertonian norms of science because this norm requires that scientific claims contribute something new, whether a new problem, a new approach, new data, a new theory, or a new explanation.

---

[6] There are other theoretical approaches that have been used for citation theories. Examples are the 'reflexive theory' by Wouters (Wouters, 1999) or the 'handicap principle' (Nicolaisen, 2007), as well as network theories (de Solla Price, 1965; Newman, 2005). We plan to explore these theories in future research.

### 3.1.2. Social constructivist theory

The focal point of this theory is that works are cited for a variety of factors, many of which have nothing to do with intellectual debt as explained by normative theory. This implies that the foundation of science originates from social actors engaging in a negotiation process in which one party convinces the other through persuasion. Thus, citations are sometimes seen as "mere persuasion" (Gilbert, 1977); accordingly, citations are merely attempts at persuading readers of the goodness of an author's claims. In essence, the social constructivist theory opposes the normative theory as it suggests that there are different motivations for citing, many of them influenced by cognitive style and personality and not necessarily by universalistic reasons. Citations are activities based on social psychological influences and are not free of personal bias or social pressures and are not always made for the same reasons. The following describes four main sources of distortion or biases: persuasion hypothesis, perfunctory citations, Matthew effect, and negational citations.

The *persuasion hypothesis* considers citations as mere 'tools of persuasion' to persuade the scientific community of the value of the work. According to White (2004), persuasion is achieved by logical arguments and inference detailed within the body of the work and by selecting important (authoritative) and adequate papers to convince readers of the importance and validity of the results resulting in a kind of 'logical' persuasion, which aligns with the universalism norm. However, White also talks about a kind of 'dark' type of persuasion that would be in line with the social constructivist theory. Within this 'dark' persuasion there are two types: 'persuasion by distortion' that occurs when "citers often misrepresent the works they allude to" (White, 2004, p. 96) and 'persuasion with names' that can be linked to the disproportionately citation of works by established authorities to gain credibility through association. *Perfunctory citations*[7], according to Murugesan and Moravcsik (1978), are citations that describe alternative approaches not utilized in the citing paper, references that are merely used to compare certain results or conclusions, references that are used to indicate the fact that a certain method employed is routine in the literature, and references that merely contribute to the chronological context of the citing paper. In other words, perfunctory citations are nonessential, superficial, redundant, or even wrong citations.

According to Merton (1968b, p. 58) the *Matthew effect* can be defined as the "accruing of greater increments of recognition for particular scientific contributions to scientists of considerable repute and withholding of such recognition from scientists who have not yet made their mark". Thus, scientists who are rich in recognition find it easier to get more recognition (and resources), which causes "the rich [to] get richer at a rate that makes the poor become relatively poorer" (p. 62). Price (1976) demonstrated the Matthew effect mathematically for publications and citations and referred to it as *cumulative advantage* or *success breeds success*, showing that the probability of being cited increases with the number of citations already obtained. This self-reinforcing effect for citations has also been shown to apply to countries (Bonitz, Bruckner, & Scharnhorst, 1997) and papers published in journals with high impact factors (Larivière & Gingras, 2010). In network theory, the Matthew effect is referred to as preferential attachment, where nodes in a network accumulate new edges proportionally to their number of edges, leading to power law distributions (Barabási, 1999; Newman, 2001). *Obliteration by incorporation* is a variant of the cumulative advantage and was also suggested by Merton (1968a), but it takes the point of view that there is an underestimation of mentions through the "obliteration of the source ideas by their incorporation in currently accepted knowledge" (Merton, 1988). In essence, papers that have become well known are not formally cited anymore. *Negational (or negative) citation* is a citation that "describes the situation when the author of the citing paper is not certain about the correctness of the cited paper" (Murugesan & Moravcsik, 1978, p. 297). In other words, these are citations to papers that may have been challenged or contradicted in other work.

---

[7] Perfunctory citations are opposed to "Organic" citations which are references to those from which concepts or theories are taken to lay the foundations of the citing paper, or papers from which certain results (including numerical ones) are taken to develop the ideas in the citing paper, or papers which help to better understand certain concepts in the citing paper (Murugesan & Moravcsik, 1978).

### 3.1.3. Concept symbols theory

The *concept symbols* theory (Small, 1978) considers the citation as symbolic of the idea expressed in the paper. The basic idea is that a citation is a symbolic act of authors associating particular ideas (i.e., concepts, procedures, or kind of data) with particular documents and is thus based on Garfield's (1964) notion of citations as descriptors in subject indexing. By using this theory one can consider citations as private symbols (cited by only one or a few authors) or standard symbols (highly cited). With a document that is repeatedly cited the citers engage in a dialogue on the document's significance, thus the meaning is conferred through this iterative activity, while at the same time the meaning of the document becomes limited through the capsulizing of a complex text into a few standard sentences (Small, 1978). This may result in the distortion or oversimplification of the original text and cause the symbolic meaning to also change over time.

## 3.2. Social theories

Researchers investigating actors and output in specific computer-mediated environments have interpreted interaction and communication using a variety of theories including theories from economics, psychology, anthropology, and sociology. In this section a select few theories will be considered in order to improve the understanding of the acts resulting in online events from which metrics are collected, these include the theories of social capital (3.2.1), attention economics (3.2.2), and impression management (3.2.3).

### 3.2.1. Social capital

Social capital is a theory that has garnered much interest as of late from a variety of disciplines. The theory stipulates that humans are social creatures and thus need to be connected to others in close-knit groups; these connections are treated as two-way investments that are maintained through reciprocal support and re-investment. Bourdieu (1985) was the first sociologist to distinguish social capital as one of three types of capital in social relations: economic, cultural, and social. Social capital can be thought of as a source of power that can be accrued through connections in a social network; actors in networks establish and maintain relationships with other actors in the hope that they may benefit in some way from these relationships. The relationships can be strong or weak (Granovetter, 1973) and this measure can have an impact on the return that an actor derives from either type including emotional support, the exchange of information, or mobilization toward a common goal. Bozionelos (2014, p. 288) used social capital theory to examine career paths in the Greek academic system and found that social capital "determines careers within that system." In social media research, several researchers (Hofer & Aubert, 2013; Steinfield, Dimicco, Ellison, & Lampe, 2009; Valenzuela, Park, & Kee, 2008) have used social capital theory to discuss aspects of interaction on various platforms.

Many other definitions in multiple disciplines have been suggested since Bourdieu's discussion of social capital (see Adler & Kwon, 2002; Portes, 1998 for summaries) including two prominent definitions from Coleman (1988, 1990) and Putnam (1995, 2000). The concept has become (in a sense) a catchall term that captures aspects of social interaction that have been studied through the lens of other concepts. Outside the arena of information communication technologies (ICT), social capital has been used to study youth behavior problems, families, schooling, public health, education, political action, community, and organizational issues such as job and career success, innovation, and supplier relations (Adler & Kwon, 2002). Social capital has become "one of the most popular exports from sociological theory into everyday language" (Portes, 1998).

### 3.2.2. Attention economics

In addition to social capital there is the theory of attention economics (Davenport & Beck, 2001), which considers the costs and benefits of finding useful information. Simon (1971) was one of the first to postulate that the world is full of information and that this takes the attention of the information consumer. When considering the theory of attention economics it is necessary to think of the growing amount of information available as a scenario in which human attention becomes increasingly valuable because there is a limited amount to be utilized. Franck (2002, p. 9) argued that scientists are "entrepreneurs who allocate time and effort so as to maximize the attention received from other scientists" and it this view that allows one to consider the ways in which scientists use tools and technologies to minimize the amount of attention they spend on sifting through the never-ending output of material to locate relevant and useful information.

Researchers have used this theoretical framework to analyze behavior within social media platforms. For example, Rui and Whinston (2011, p. 322) examined approximately 3 million Twitter users and found that social media environments are a "marketplace where people contribute information to attract attention and contribute attention while consuming information." The attention economy framework has also been used to evaluate novelty and popularity in social networks (Huberman, 2013) and to examine pedagogical strategies for retaining the attention of law students in the technology-rich environment of today's classroom (Matthews, 2012). The attention level of an audience member is determined by their "attention capacities and on the total volume of signals to which they are exposed" (Falkinger, 2003, p. 4), and today's environment exposes scholars to an unconscionable amount of information suggesting that it is extremely important to consider how they manage and conserve attention.

### 3.2.3. Impression management

Finally there is a dramaturgical framework put forth by Erving Goffman (1959) in which he describes activities called self-presentation and impression management. Impression management is a process that takes place as humans interact with one another and is motivated by the need to avoid shame and embarrassment, while self-presentation is the act of presenting information about oneself to an audience. Goffman described these processes using dramaturgical concepts that include actors, audience, and stage, and wrote that when people interact with one another they act out a role for their audience and are required to maintain the impression of that role through the entirety of the interaction; if the impression is broken and the audience loses faith in the presentation of the role, the actor will be shamed and embarrassed.

The concepts of impression management and self-presentation have been defined in the literature many ways, with most building upon Goffman's description. Gosling and colleagues (Gosling, Gaddis, & Vazire, 2007) examined the accuracy of impressions in Facebook finding that personality impressions were limited in accuracy and that authors did enhance their own self-presentations. When comparing impressions made on Facebook with impressions made in face-to-face meetings, Weisbuch and colleagues (2009) found that they were very similar. Gilpin (2011, p. 234) writes that tweeting plays an important role in impression formation, "as followers will primarily draw conclusions based on the contents of tweet messages as well as indications of the intended recipients of those messages."

## 4. Applying theories to selected acts

Using the framework in Figure 1 to describe, define, and distinguish various acts related to research objects, we discuss the citation and social theories introduced in section 3. Due to space restrictions, this work will focus solely on applying these theories to acts related to the scientific journal article[8] as the most common and important type of scholarly documents (and the focus of most currently captured altmetrics). Future work will consider discussing acts related to other types of documents as well as scholarly agents. In order to simplify the discussion and provide more room for detail, the focus will be on some of the most popular acts in terms of number of captured online events to scholarly papers as well as researched events in the current altmetrics setting[9]. These include the acts of *saved in Mendeley* (4.1.1) for the *access* category, *mentioned in a tweet* (4.1.2) and *reviewed on F1000* (4.1.3) for *appraise,* and *cited in a blog post* (4.1.4) for a specific case of *applying* the content of a scientific journal paper. Mendeley reader counts and tweets have been shown to be the most prevalent online events currently captured for scientific papers[10]; Mendeley reader counts account for two thirds of recent journal articles and tweets mention approximately one fifth of recent journal articles. Reviews on F1000 and blog citations occur much less due to their selectivity and higher level of engagement, but are discussed as they represent particular forms of acts regarding journal articles (Bornmann, 2014b; Costas, Zahedi, & Wouters, 2014; Haustein, Costas, & Larivière, 2015; Haustein, Larivière, et al., 2014).

### 4.1. Access: Saved in Mendeley

A Mendeley readership count for a particular document (at this time) indicates that a Mendeley user has added the document to his or her Mendeley library[11]. Users of Mendeley are assumed to have an interest in organizing bibliographic metadata to keep track of and manage scientific documents either for citing or using them in a professional or educational context (which could include teaching and self-teaching). However, each document added to a Mendeley user library is not necessarily read (Mohammadi, 2014) and there is no guarantee that a user adds all documents he or she has (or intends to) read or cited to their Mendeley library.

According to previous research (Mohammadi, Thelwall, Haustein, & Larivière, 2014; Mohammadi, 2014), most Mendeley users are students, postdocs, and researchers and as such it is assumed that Mendeley readership counts are a reflection of interest by a scholarly audience beyond the community of citing authors. It is not known whether the approximately three million Mendeley users (Haustein & Larivière, 2014) are representative of the entire readership of scientific documents and whether certain biases exist regarding disciplines, academic age, and countries. Among the currently captured social media metrics related to scientific documents, Mendeley reader counts have the highest correlations with citations, ranging from medium to high values (e.g., Mohammadi et al., 2014; Mohammadi & Thelwall, 2014; Zahedi, Costas, & Wouters, 2014), which implies a certain similarity between the two metrics. This suggests that citation theories may be of value to understand what is happening in the Mendeley environment.

---

[8] In this context the journal article as the research object may also refer to other versions of the original publication such as the preprint/eprint on repositories, which may or may not be considered as one single research object.

[9] See for example Bornmann (n.d.); Costas, Zahedi, & Wouters (2014), Haustein, Costas, & Larivière (n.d.), Haustein et al.( 2014), Holmberg & Thelwall (2014), Mohammadi & Thelwall (2013) and Waltman & Costas (2014).

[10] It should be noted that due to current technical constraints, the scholarly document being tracked is (in most cases) a peer-reviewed journal article with a DOI or comparable identifier (e.g., arXiv id or PMID).

[11] Technically, a user can save a document either online or in the desktop version of the reference manager. Bibliographic information is typically extracted directly from the document or the metadata is harvested from the website where the document is located. This may lead to errors in the bibliographic data that can be manually corrected by the user.

Opposed to citing, where the different norms of communism, universalism, disinterestedness, and organized skepticism (and originality) are expected to apply, Mendeley users may not necessarily adhere to these norms when adding documents to their libraries – in fact, norms regarding literature management do not exist. Students and young scientists presumably do not (yet) learn that they are supposed to save and organize all relevant documents in reference management software. When saving papers, the principle of "giving credit where credit is due" does not necessarily apply because documents are often added before they are actually read, which implies that Mendeley user libraries do not only include the most influential and relevant documents, but also include those without actual value to the user. The act of saving a document to Mendeley is assumed to be more general than the act of citing, because more documents are saved and read than cited. Moreover, saving to Mendeley may not only be due to utilitarian reasons of saving and organizing, intended reading, highlighting, and annotating, which would imply a certain level of giving credit to authors, but also due to (self-)marketing. However, if the act of *saving to Mendeley* is considered *use* of the document—including various facets from mere saving to intense reading, annotating, and citing—then Merton's communism, as well as universalism, could apply because authors receive Mendeley reader counts when their papers are saved to Mendeley. If users added all documents they read to their Mendeley libraries and had read all documents in their Mendeley libraries, Mendeley readership would give credit to the sources their knowledge is based on without distinguishing to what extent these documents were relevant or required to fulfill the library owner's information needs. In this regard, the act of saving to Mendeley as reflected in reader counts could be considered as a pellet of peer recognition similar to citations, although certainly on a different level of engagement and without any quality control or space restrictions[12] of peer-reviewed journal articles. However, a survey among 679 Mendeley users found that only 27% of users had read all of the documents in their libraries (Mohammadi, 2014). It can be argued that the disinterestedness norm applies to saving to Mendeley, but more as an unconscious act given that saving to Mendeley is essentially anonymous[13]; this is in contrast to the act of citing, which can be considered "a private process with a public face" (Cronin, 1981, p. 16). The norm of organized skepticism does not apply to the act of saving to Mendeley because documents are often added before they are read, thus they are added without being considered 'skeptically'. Organized skepticism could only apply if a document was scrutinized by the user and then removed from their Mendeley library.

Empirical user studies have yet to show whether and to what extent users adhere to norms when saving documents to Mendeley in order to determine whether Mendeley reader counts might signify influence (albeit in a broader sense than citations). Even if the normative theory does not yet apply to the act of saving to Mendeley due to the current lack of equivalent norms regarding the use of reference managers, norms regarding literature management could be introduced in the future to establish *saving to Mendeley* (or any other reference manager) as an inherent part of the scholarly communication process.

---

[12] Note that technical limitations such as available space in user libraries could still apply.
[13] It should be note that the link between user and document is visible on other platforms (e.g., CiteULike).

The value of the social constructivist theory to interpret this event would stem from its ability to interpret the act in terms of its pre-citation role and, indeed, a survey among Mendeley users found that the main reason to save documents to Mendeley was to cite them (Mohammadi, 2014). It is, however, difficult to expect that someone would simply save a document in Mendeley for persuasive or perfunctory reasons given the anonymous nature of the act on Mendeley. Saving a document as a negative example would conceptually be possible, although identifying this kind of negative use is currently not possible.

The Matthew effect could apply to Mendeley in a manner similar to citations. Considering the cumulative advantage within the platform, documents that have already been saved to Mendeley libraries are more likely to be added by other users because they appear in Mendeley search results and when browsing the Mendeley website. Applying the Matthew effect in a more general sense, considering various aspects of scientific and social capital, Mendeley users would be prone to save more documents from renowned authors and high impact journals. Findings by Costas, Zahedi and Wouters (Costas, Zahedi, & Wouters, 2015) note that articles published in high-impact journals such as *Nature* or *Science* account for a substantial amount of reader counts suggesting that some kind of Matthew effect applies when documents are saved to Mendeley.

Apart from the pre-citation context—saving a document to Mendeley in order to cite it—the idea of concept symbols could be applied to tagging or summarizing a document in Mendeley and thus make it a particular symbol for the Mendeley user. However, this would be more in the line of appraising than accessing a Mendeley document.

Using the theory of social capital to interpret why a user saves a document in Mendeley can also be useful. Users may save publications into the system with hope that it will increase their worth by increasing the visibility of their own work. As social network connections are seen to be of value, it makes sense that increasing the visibility of one's work has the potential of increasing one's social capital in that network.

When examining this act from the perspective of attention economics, again this social theory is well suited to explain why a user saves documents in Mendeley. He or she is making use of Mendeley (such as searching, storage, and organization) to reduce the amount of attention they will need to utilize it in the future. Saving a publication to Mendeley allows the user to reduce the amount of time and effort they spend on information sifting so that they may spend their valuable attention on other matters.

If one applies impression management theory to the act of saving a document in Mendeley, it becomes clear that a user may save their own publication to Mendeley (and thus the publications appear in the Mendeley search or when browsing the Mendeley website) in order to impress upon others that they are accomplished in their area of study or that they are merely meeting certain impressions others (such as their colleagues, students, or administrators) have of them.

### 4.2. Appraise: Mentioned in a tweet

Due to the 140-character limitation of a tweet, a scientific document has to be referred to by a commonly used unique identifier or URL[14]. Perhaps the most important difference between citations and tweets is that the former is a standardized and codified type of mention, while in Twitter the norms surrounding the mentions are rather different. Many documents are highly tweeted not due to their scientific merits (Haustein, Peters, Sugimoto, Thelwall, & Larivière, 2014), but instead often reflect "ephemeral (or prurient) interest, the usual trilogy of sex, drugs, and rock and roll" (Neylon, 2014, para. 6), which discredits Merton's notion of valuing knowledge claims. Twitter users do not seem to (and they are actually not expected to) concern themselves with whether or not the document is original, if it's of high quality, if they are rewarding the authors, etc., instead the Twitter environment seems to be mostly free of these expectations (or at least not intrinsically based on the Mertonian norms). It could still be argued that some degree of communism, universalism, or disinterestedness apply to a specific subset of tweets, if Twitters use scientific publications to discuss, debate, or contrast scientific ideas on Twitter. Findings regarding bot accounts, which automatically tweet links to scientific documents, are a another example of the limitations of Mertonian norms to this type of act (Haustein, Bowman, et al., 2014), as automated diffusion cannot be considered a social act.

Finding aspects of social constructivism in tweets mentioning scientific publications is hampered by the 140-character restriction. However, it can be argued that some forms of persuasion might be attributed to Twitter users when mentioning a scientific document. Perfunctory tweets, for example wrongful mentions of papers linked to unrelated topics or presenting authoritative references for invalid arguments, are possible. However, these might instead be considered misunderstandings or 'misframings' (Goffman, 1974) by their users instead of a conscious act of manipulation of the counts of publications with superficial or wrong mentions. The findings that retracted publications receive more Twitter mentions than regular papers (Haustein et al., n.d.) supports the idea that negative mentions of scientific papers on Twitter occur, but a small case study (Thelwall, Tsou, Weingart, Holmberg, & Haustein, 2013) suggests that they might be as rare as negative citations in the natural sciences and medicine (Murugesan & Moravcsik, 1978).

The Matthew effect might play an important role in the accumulation of tweets for scientific documents, for example within the platform itself through affordances like the retweet function. Thus a (re)tweeted paper would increase its Twitter visibility and accrue more (re)tweets as an effect of its previous (re)tweets. Twitter users receive push notifications if a particular number of users they follow has retweeted the same tweet (Satuluri, 2013), which might produce further tweets. Another Twitter-specific aspect related to the cumulative advantage is the number of followers. A paper mentioned by an account with a large number of followers can be expected to be visible to a larger audience, thus increasing its potential to receive more tweets. This might, for example, be the case when official Twitter accounts of scientific journals tweet their papers (Haustein, Peters, et al., 2014). The Matthew effect can be also argued to apply in a more general sense; for example, when a paper is frequently mentioned on Twitter due to the scientific capital of the authors or the journal it was published in. It is quite likely that success on Twitter is bred by a mix of social capital within the platform—number of (re)tweets and followers—as well as in the scientific community (e.g. reflected by the citation impact or funding success or even awards). This is supported by the finding that the highest number of tweets are obtained by papers published in journals such as *Nature*, *Science*, *New England Journal of Medicine* and *Lancet*, which belong to the most prestigious scientific journals and, at the same time, promote their content through official Twitter accounts with many followers (Costas et al., 2015; Haustein, Peters, et al., 2014).

---

[14] Mentions of scientific papers on Twitter are currently captured if they include the publisher URL or common document identifiers such as DOI, PMID or arXiv ID. Formal or informal citations such as *Cronin, B. (1981). The need for a theory of citing. Journal of Documentation, 37(1), 16–24* or *Cronin's paper in JDoc* are not captured.

The fast and brief nature of tweeting promotes obliteration by incorporation in so far as users might avoid linking to the original paper due to space limitations. This could end up leading to an under valuation of certain documents.

The concept symbols theory has a special fit for the act of tweeting about scientific papers. The short nature of tweets strongly supports the narrowing of meaning stipulated by this theory, as the meaning of (or engagement of the Twitter users with) the publications has to be encapsulated in 140 characters. Also, the association of publications with hashtags (which can be signs or signifiers referring to ideas or concepts) supports the idea that mentions of documents on Twitter are concept symbols. Similarly, when a document is repeatedly tweeted and Twitter users engage in a discussion about the value of the document, the meaning of the tweets is conferred through the symbolic usage of the publications (similar to the use of citations as symbols). This narrowing of meaning and the fact that many of the users are not necessarily experts on the paper's topic may result in the distortion or oversimplification of the original text (similar to citations), a phenomena that has been already discussed for altmetrics (Colquhoun, 2014).

Approaching the act of mentioning in Twitter using social theory can provide additional insight. Users may have many different motivations for mentioning a scientific document in a tweet, yet the theory of social capital suggests that one of these motivations will be to establish a connection between the tweeter and the publication (and in return the author(s) of the document). When a scholar tweets about a scientific document, they are making a weak connection in the network between themselves and the other. If the tweeter continues to tweet about publications from the same author(s), the connection between the two (or more) can become stronger creating a potential form of revenue that can be later converted into benefits such as a collaboration or a letter of reference.

When examining the Twitter mention from the perspective of attention economics, the tweeter will use Twitter-specific affordances (such as the URL, an @ symbol, and/or a #hashtag) to reduce the amount of attention they and others will need to access and understand the paper, as the searching and organization in the Twitter environment is facilitated by the use of these affordances.

Looking at this act through the lens of impression management allows one to interpret the act as an attempt to impress upon others that he or she is up to date on the work in their field, that they present themselves as one who can be trusted to spread relevant documents, or that they are simply reinforcing their identity as an expert in a specific domain.

### 4.3. Appraise: Reviewed on F1000Prime

Faculty of 1000 (F1000Prime or F1000[15]) is a commercial online post-publication peer review and recommendation service for biological and medical research launched in 2002. More than 5,000 peer-nominated researchers and clinicians, referred to as F1000 faculty members, produce the reviews. Faculty members are requested to select the most interesting publications they read and to provide reviews of these publications. A review of a publication consists of a recommendation ("good," "very good," or "exceptional") along with an explanation of the strengths (and possibly also the weaknesses) of the publication. Faculty members can choose to review any primary research article from any journal without being limited to recent publications or publications indexed in any given database (Waltman & Costas, 2014). Papers are not only rated in F1000, but they are also reviewed and labeled in order to indicate such things as the appropriateness for changes to clinical practice, suitability for new drugs, or usability for teaching (Mohammadi & Thelwall, 2013).

---

[15] See http://f1000.com/prime

Given the scholarly nature of F1000 reviews the normative theory can apply in a similar manner to citations. F1000 faculty are expected to behave according to the 'ethos of science' when selecting and reviewing their publications for F1000, as "they must sign a statement to indicate that the article has been selected for inclusion in F1000Prime's Article Recommendations entirely on its scientific merit and that they have not been influenced directly or indirectly in the selection of articles by the authors or any third party"[16]. The F1000 post-publication peer review system shares advantages as well as some disadvantages and biases—e.g. subjectivity, lack of consensus and biases of referees, cost and time issues, and difficulty to scale (e.g., King, 1987; Lee, Sugimoto, Zhang, & Cronin, 2013)—with traditional closed blind peer review, which makes the normative theory highly applicable. In addition to the traditional pre-publication review, recommendations on F1000 are attributed and linked to referees and there are various steps of quality control, such as monitoring by section heads and the possibility for faculty members to disagree with recommendations[17], which should further encourage referees to behave according to Mertonian norms.

For the same reasons, persuasive and perfunctory reviews and recommendations are quite unlikely and negative reviews are technically not possible on F1000[18], although constructive criticism is encouraged[19]. Although publications from all types of sources and authors are expected to be reviewed, the higher accumulation of reviews in high impact journals (Waltman & Costas, 2014) points to a form of the 'Matthew effect' in the selection and review of documents by F1000 similar to what Larivière and Gingras (Larivière & Gingras, 2010) found for citations. The cumulative advantage in a stricter sense, in that papers with many F1000 recommendations are likely to accrue even more within the same platform, is less likely to apply because F1000 is supposed to function as a filter where experts choose the most important articles from their research area. This assumption is supported by findings that the vast majority of reviewed publications have only one review and less than 2.5% receive more than 3 recommendations (Waltman & Costas, 2014).

The concept symbol theory applies in so far that faculty members choose and recommend documents for particular reasons that are indicated by particular tags, such as *new findings*, *controversial*, or *good for teaching* or in the recommendation text in the form of concrete statements about the contents, interest, and usefulness of the documents. A recommended document could thus be considered a private symbol of the reviewer, who distills the meaning of the original text in his or her review. The idea of standard symbols is less likely to apply, as most documents are only recommended by one faculty member.

When a scholar reviews an article he or she is establishing a (weak) connection between his or her and the review itself, and subsequently the venue in which the review occurs, the content of the article, and potentially the author(s) of the document. This connection can be of value to the reviewer at a later time as they can list this service on their CV or they can help establish trends in research areas by reviewing (what they consider) quality documents, thus supporting the interest of the social capital theory in the understanding of F1000 recommendations.

---

[16] http://f1000.com/prime/about/whatis
[17] http://f1000.com/prime/about/whatis
[18] It should be noted that negative reviews are possible on other platforms (e.g. Publons or Pubpeer).
[19] http://f1000.com/prime/about/whatis

From another perspective, attention economics might suggest that a reviewer would choose to participate in this act to reduce the amount of information they need to search by focusing on documents relevant to their own work. This in turn would reduce the amount of documents they would need to examine when trying to find information in the future.

If one were to use impression management to examine this act, it is clear that the reviewer may be looking to create a new impression of himself or herself as someone who can be relied upon to act as a gatekeeper of science or to reinforce an existing impression. The affordances of the platform allows for the establishment of a reviewer as a "global expert"[20] and this then allows them to present a self that contributes to the impression others have of them.

### 4.4. Apply: Cited in a blog post

Among current social media activity, the act of citing in blogs is believed to be the most similar to citing in a scientific documents, because research blogs allow for and provide the space to discuss and analyze scientific content (Shema & Bar-ilan, 2014). The findings that blogs have moderate correlations with citations (Costas et al., 2014; Haustein et al., n.d.; Shema, Bar-ilan, & Thelwall, 2014) support this assumption. As a result, citation theories are more likely to apply to blog citations than mentions on Twitter. However, even in the case of *applying* a scientific document to a blog (e.g. building upon and reusing results, methods and theories of a scientific documents[21]) particular differences between scientific peer-reviewed documents and blog posts limit the applicability of citation theory; this can be caused by blogs not having the same scholarly nature as scientific publications as they lack the gatekeeping and quality control present in academic work. Anyone can publish a blog on their homepage, posts are typically not peer-reviewed, and some blog posts simply announce the publication of interesting articles. It should be noted, however, that some blog aggregator platforms (such as ResearchBlogging.org) have bloggers agree to specific guidelines[22] that ensure that the posts of bloggers on the platform discuss peer-reviewed research which they have fully read, understood, and formally cited, while adding original content.

The normative theory could thus be applied to mentions in blogs, if one would expect bloggers and science journalists to cite according to similar 'norms' as scholars. However, given the more open and less controlled nature of blogs, it is reasonable to think that the application of these rules would be less strict than in scholarly contexts.

From another perspective, persuasion is indeed a driving force in blogging and scientific journalism by frequently discussing scholarly information and presenting the viewpoints of their authors on scientific issues (Shema, Bar-Ilan, & Thelwall, 2012; Shema & Bar-ilan, 2014). However, it could be argued whether the value of name-dropping and perfunctory mentions have the same incidence in blog posts as in scholarly publications. Considering that bloggers and science journalists do not share the same reward system as scholars, it is arguable that the presence of these types of persuasions may be less frequent in this environment, although the possibility still exists that other types of rewards (e.g. followers, visitors of their blogs, commenters, etc.) may play a role in influencing bloggers' behaviors (e.g. by distorting the content of scientific documents in order to support their views or to mention publication from esteemed journals or authors). The Matthew effect is also applicable for blog posts, as it is plausible that bloggers and scientific journalists primarily focus on well-established, famous, and popular authors or journals (Shema et al., 2012). Finally, the existence of important blogging activities around retracted publications supports the idea that negative mentions are also important in the consideration of blog mentions.

---

[20] As indicated on the F1000 website (http://f1000.com/about-and-contact)
[21] This seems plausible, as reported by Fausto and colleagues (Fausto et al., 2012) the number of citations per blog is increasing but with values of citations per blog post between 1.38 to 1.48
[22] http://researchblogging.org/news/?p=53

The concept symbols theory can be applied to blog mentions in a similar manner as to citations; being reinforced by the open-natured and laymen-authored activity of typical blogs, which would allow for the stronger narrowing down of the document's original meaning.

Social capital, attention economics, and impression management all provide useful insight into why a blogger or journalist may cite a scholarly document in their blog. For instance, the blogger establishes a weak connection between her blog and the scholarly document (and subsequently the author(s) and/or the journal) that can become stronger by the continued use of other documents by the same author(s), thus providing social capital to the blogger by creating a connection between her and the author(s) of the documents, or by simply driving traffic to her blog.

The creation of blogs around scientific documents also allows the blogger (and her users) to keep a record of 'useful' (or not useful) academic work that can be easily filtered through the typical blogging affordances of searching, tagging, and storing posts, thus decreasing the attention needed to find these documents in the future.

Finally, impression management may explain why a blogger might cite a specific scholarly document, as the blogger must be concerned about the impression he or she is creating by applying the document. Thus the blogger may be attempting to create an impression that implies they understand and have expertise on the cited material.

## 5. Conclusions and Outlook

In the current debate around so-called altmetrics, some argue that these *polymorphous mentions* (Cronin et al., 1998, p. 1320) can be a good proxy for societal impact (Bornmann, 2014b), early scientific impact (Eysenbach, 2011), attention, and educational and practical use (Mohammadi et al., 2014; Zahedi, Costas, & Wouters, 2013), while others argue that they reflect nothing but buzz, popularity, or simply increasing visibility (Colquhoun, 2014). The answer is probably that the new metrics are all of the above and the extent to which each of these occurs depends the particular platform, its uptake and users, as well as on the research topic, the unit of analysis, and the context of the metric. In general, what is still lacking is a concrete set of frameworks, models, and theories that can help to support and frame interpretations and uses of these new social media metrics.

In this chapter a first step has been taken towards providing a conceptual framework that can be applied to the acts leading to (online) events underlying these new metrics in the context of scholarly communication. In order to better comprehend these heterogeneous acts, the framework identifies three broad categories of acts related to scholarly documents and agents: accessing, appraising, and applying.

Common citation theories and social theories are introduced to discuss whether they can be used to explain the acts underlying the various indicators referred to as altmetrics. The reason to begin theoretical discussions by applying normative, social constructivist, and concept symbols theories to social media metrics is based on the strong, albeit antagonistic, relationship of altmetrics to citations (Priem et al., 2010; Priem, 2014). In addition to these citation theories, three social theories were used to interpret these events including social capital, attention economics, and impression management because of the inherent social nature of these platforms. Focusing on the journal articles as the most common form of scholarly documents, these theories were applied to the acts of saving to Mendeley, mentioning on Twitter, reviewing on F1000 and citing in a blog post in order to determine their applicability.

Due to the heterogeneity of various types and categories of acts (i.e., access, mention, appraise, and apply), the discussed citation theories are more or less suitable. Similar to Hogan & Sweeney (2013) who discussed the limited fitting of Mertonian norms to the new social media, this work found that Mertonian norms fit quite well with the act of reviewing and recommending on F1000, to a lesser extent with being cited in blogs, and that the norms did not really fit with being mentioned on Twitter. In the context of saving to Mendeley, the normative theory is more likely to apply only in the pre-citation context, i.e. when it is linked to the act of citing. Indeed, 85% of Mendeley users surveyed by Mohammadi (2014) claimed that they saved documents to the reference manager in order to cite them. However, only 27% stated that they had read all of the documents in their user libraries.

In terms of social constructivist theories, the Matthew effect idea has the strongest potential for most of the discussed acts and it can, to a certain extent, explain the concentration and skewness of social media events across publications. This might be attributed to the networked nature of these platforms. Thus, platforms such as Mendeley and Twitter support the necessary processes most often associated with the Matthew effect and preferential attachment, as documents with more events get higher visibility within the platforms through different mechanisms (e.g. re-tweets, number of followers or Mendeley filtering tools). For F1000 reviews, as well as blogs, the higher presence of papers from prestigious journals suggests that the Matthew effect could apply from the point of view of the concentration of events around specific agents (e.g. prominent authors, journals, etc.) represented in the system. The presence of the Matthew effect may have important implications for the potential consideration of acts in social media metrics when considering the reward and communication system of science.

Concept symbols are more likely to apply to the act of tweeting documents in so far as Twitter could serve as a language system with particular symbols (e.g., hashtags linked to publications) indicating a particular idea or concept in relation to the document. The applicability of this theory would conceptually support the notion of tweeted papers being used as concept symbols by an audience broader than the scientific community. Thus, it would be reasonable to use Twitter as a tool to capture the public perception of science, particularly if these concepts regarding scientific documents differ from those of the scientific community[23]. To a lesser extent than Twitter, the theory of concept symbols could also be applied to the acts of blog mentions and F1000 reviews in a similar fashion as to citations.

---

[23] For example, a common criticism is that Twitter users do not understand scientific publications and therefore they mention them for other reasons than their scientific merit. Identifying these different symbolic connections generated by the general public would help to develop mechanisms to improve the understanding of research results outside the scientific community.

In addition to citation theories, three social theories were used to interpret these acts: social capital, attention economics, and impression management. Each of these theoretical viewpoints allows one to interpret the specific acts described above using different lenses. From a purely social perspective, social capital explains how the use of these platforms benefits the scholars by providing them with a network of potential resources to mine and utilize when necessary. From a pragmatic view, attention economy is useful as it discusses the benefits of using social media to decrease the time spent both finding and attending to information sources. Finally, impression management describes the ways in which scholars must actively maintain their presentation of self as they navigate the blurring boundaries of the public/private nature of social media.

It is important to keep in mind that these new metrics may be influenced by 'noise' (i.e. all mentions that are not meaningful or deviant from the intended meanings such as 'automated' mentions, self-mentions, data errors, etc.) and that this noise can introduce doubt as to how to properly interpret the significance of the acts leading to these captured metrics. In addition, the technology and its affordances are constantly changing so that the acts themselves are being affected, which in turn can bring new challenges and issues to the understanding of these metrics.

Given the heterogeneity of the acts, the theories discussed in these chapter cannot fully explain social media acts related to scholarly communication and empirical research – such as content analyses along the lines of Shema et al. (2014), as well as user surveys like Mohammadi (2014) – is needed to further investigate user motivations behind these acts to increase the understanding of these various metrics and validate their use in research evaluation.

Mohammadi, E., & Thelwall, M. (2014). Mendeley readership altmetrics for the social sciences and humanities: Research evaluation and knowledge flows. *Journal of the Association for Information Science and Technology*. doi:10.1002/asi.23071

Mohammadi, E., Thelwall, M., Haustein, S., & Larivière, V. (2014). Who reads research articles? An altmetrics analysis of Mendeley user categories. *Journal of the Association for Information Science and Technology*. Retrieved May 21, 2014, from http://www.scit.wlv.ac.uk/~cm1993/papers/WhoReadsResearchArticlesPreprint.pdf

Murugesan, P., & Moravcsik, M. J. (1978). Variation of the nature of citation measures with journals and scientific specialties. *Journal of the American Society for Information Science*, *29*(3), 141–147. doi:10.1002/asi.4630290307

Newman, M. E. J. (2001). Clustering and preferential attachment in growing networks. *Physical Review E*, *64*(2), 025102. doi:10.1103/PhysRevE.64.025102

Newman, M. E. J. (2005). Power laws, Pareto distributions and Zipf's law. *Contemporary Physics*, *46*(5), 323–351.

Neylon, C. (2014). Altmetrics: What are they good for? *PLOS Opens*. Retrieved from http://blogs.plos.org/opens/2014/10/03/altmetrics-what-are-they-good-for/

Nicolaisen, J. (2007). Citation Analysis. *Annual Review of Information Science and Technology*, *41*, 609–641.

Otlet, P. (1934). *Traité de documentation. Le Livre sur le Livre: Théorie et Pratique* (Editiones .). Bruxelles: D. Van Keerberghen & fils.

Piwowar, H. A. (2013). Altmetrics: Value all research products. *Nature*, *493*(7431), 159. doi:10.1038/493159a

Portes, A. (1998). Social capital: Its origins and applications in modern Sociology. *Annual Review of Sociology*, *24*, 1–24.

Price, D. D. S. (1976). A general theory of bibliometric and other cumulative advantage processes. *Journal of the American Society for Information Science*, *27*(5), 292–306. doi:10.1002/asi.4630270505

Priem, J. (2010). I like the term #articlelevelmetrics, but it fails to imply *diversity* of measures. Lately, I'm liking #altmetrics. *21 September 2010, 3:28 a.m. Tweet*.

Priem, J. (2014). Altmetrics. In B. Cronin & C. R. Sugimoto (Eds.), *Beyond bibliometrics: harnessing multi-dimensional indicators of performance* (pp. 263–287). Cambridge, MA: MIT Press.

Priem, J., Taraborelli, D., Groth, P., & Neylon, C. (2010). Alt-metrics: a manifesto. *October*. Retrieved from http://altmetrics.org/manifesto/

Putnam, R. D. (1995). Bowling alone: America's declining social capital. *Journal of Democracy*, *6*(1), 65–78.

Putnam, R. D. (2000). *Bowling Alone: The Collapse and Revival of American Community*. New York: Simon & Schuster.

Riviera, E. (2014). Testing the strength of the normative approach in citation theory through relational bibliometrics: The case of Italian sociology. *Journal of the Association for Information Science and Technology*. doi:10.1002/asi.23248

Ronald, R., & Fred, Y. Y. (2013). A multi-metric approach for research evaluation. *Chinese Science Bulletin*, 10–12. doi:10.1007/s11434-013-5939-3

Rui, H., & Whinston, A. (2011). Information or attention? An empirical study of user contribution on Twitter. *Information Systems and E-Business Management*, *10*(3), 309–324. doi:10.1007/s10257-011-0164-6

Satuluri, V. (2013). Stay in the know. *Twitter blog post 24 September 2013*. Retrieved from https://blog.twitter.com/2013/stay-in-the-know

Shema, H., Bar-ilan, J., & Thelwall, M. (2014). Do blog citations correlate with a higher number of future citations ? Research blogs as a potential sources for alternative metrics. *Journal of the Association for Information Science and Technology*, *65*(5), 1018–1027. doi:10.1002/asi.23037

Shema, H., Bar-Ilan, J., & Thelwall, M. (2014). How is research blogged? A content analysis approach. *Journal of the Association for Information Science and Technology*. doi:10.1002/asi.23239

Simon, H. A. (1971). Designing organizations for an information-rich world. In M. Greenberger (Ed.), *Computers, Communications, and the Public Interest* (pp. 37–72). John Hopkins Press.

Small, H. (1978). Cited documents as concept symbols. *Social Studies of Science*, *8*, 327–340. doi:10.1177/030631277800800305

Small, H. (2004). On the shoulders of Robert Merton: Towards a normative theory of citation. *Scientometrics*, *60*(1), 71–79. doi:10.1023/B:SCIE.0000027310.68393.bc

Steinfield, C., Dimicco, J. M., Ellison, N. B., & Lampe, C. (2009). Bowling Online : Social Networking and Social Capital within the Organization. In *4th International Conference on Communities and Technologies* (pp. 245–254). University Park, Pennsylvania: ACM Press.
23